# A Joint Calibration Technique for Improving Measurement Accuracy of Voltage and Current Probes During Synchronous Operation for RF Based Plasma Devices

Mahreen Khan[1], A. Ganguli[2], Veda Prakash Gajula[3], Satyananda Kar[4*], Debaprasad Sahu[5]

Department of Energy Science and Engineering, Indian Institute of Technology Delhi,

Hauz Khas, New Delhi – 110016, India

[1]mahreen.khanm@gmail.com,    [2]ashishganguli@gmail.com,    [3]prakashgveda@gmail.com, [4]satyananda@dese.iitd.ac.in, [5]dpsahu@dese.iitd.ac.in

*The author to whom correspondence may be addressed: satyananda@dese.iitd.ac.in

**ABSTRACT**

This paper presents a *joint calibration* scheme for voltage (V) and current (I) probes that helps resolve accurately voltage-current phase differences even when the difference is very close to 90°. The latter has been a major issue with V-I probes when used with miniature RF plasma devices like the atmospheric pressure plasma jet (APPJ). Since the impedance of such miniature devices is predominantly capacitive, the phase difference between the voltage and current signals is very nearly 90°. It turns out, however, that when V-I probes are used with such devices without joint calibration, these frequently yield phase shifts over 90°. Also, since power absorption is proportional to the *resistive* part of the impedance it becomes very sensitive to the phase difference when it is close to ≈ 90°. Thus, it is important to be able to resolve the phases accurately. Post-calibration, V-I probes would be indispensable for the electrical characterization of APPJs for determining average RF power $P_{av}$, plasma impedance $Z_p$, etc. Typical post-calibration V-I data yields, $Z_p \approx 93.6 - j\,1139\ \Omega$ (81.5 - j 1173 Ω) at $P_{av} \approx 9.8$ W (≈ 7.7 W) for helium (argon) gas.

## I. INTRODUCTION

Voltage (V) and Current (I) probes are very common measurement tools that are used over a wide range of frequencies. In particular, these are used extensively for RF-based plasma devices like Capacitively Coupled Plasmas (CCPs), Atmospheric Pressure Plasma Jets (APPJs), etc, where the probes are used to measure the complex voltage and current being fed to the device.[1–7] When used for such applications, the V-I probes are inserted between the matching network (MN) and the plasma load (PL), preferably close to the load.[2,8–11] A serious problem that plagues the V-I method at radio frequencies, is that for large capacitive impedances, where the phase difference between the current and voltage is nearly 90°, the *measured phase difference can actually exceed* 90°, giving a negative plasma resistance and negative power coupling to the plasma. The problem is particularly severe for miniature devices like RF APPJs, where on account of the small plasma sheath and glass capillary capacitances that come in series with the plasma, one obtains a very large capacitive impedance.[1,3,6,7,10,12–14] In this context, one may note that the characterization of RF APPJs has remained an important issue even today due to their small size and the high cost of advanced diagnostic tools, [6,15–23] although such devices are rapidly proving to be highly suitable for a wide range of applications.[24–28] Under such circumstances, V-I probes being relatively cheap are the natural choice for measuring the electrical parameters of the plasma. In fact, one *expects* that when placed between the MN and the PL, these *should* be able to provide highly reliable measurements of the complex voltage and current fed to the plasma (and hence the real and reactive power flow), the forward / reflected wave amplitudes on the line, the plasma impedance, the harmonics produced by the *nonlinearity* of the plasma sheath, etc. It turns out, however, that while this is possible at low frequencies, at radio frequencies matters can become complicated on account of high frequency and transmission line effects (and difficulties due to RF interference).[1,6,8,10,29]

It may be noted that the latter problem is not related to the frequency response of the (V- I) probes themselves. It turns out that straightforward use of even high-quality probes for RF plasma jets yields the absurd results cited above. The problem arises since most often commercially procured voltage and current probes are only provided with *the signal magnitude or scale conversion data* with no corrections for time or phase delays *between the probe output read on the oscilloscope and the true signal value at the location of the probe*. Such data is particularly crucial for *high bandwidth* probes since it is eminently possible for time or phase errors to arise in such probes on account of *variations in the time delays in probe electronics and in the electrical lengths of the cables with the frequency*, particularly at radio frequencies.

In the literature, one finds several attempts to resolve the problem plaguing the use of V-I probes for APPJs applications.[1,3,6,7,10,29] However, these efforts seem somewhat ad hoc and makeshift rather than based on sound RF techniques, and to the best of the authors' knowledge/understanding, such efforts have been unable to resolve the problem satisfactorily till now. It may be mentioned that dual directional couplers (DDCs) placed between the matching network (MN) and the RF generator have also been used.[29] However, in the latter configuration, the DDC cannot yield any information on the forward/ reflected waves, plasma impedance, harmonics produced by the plasma, etc. Moreover, losses in circuit components render the actual power fed to the plasma device uncertain. In order that a DDC can effectively replace the V and I probes, it must be placed *between the MN and the plasma load.*[30] As shown by Rawat *et al,* [29] when used this way DDCs can furnish all relevant information although their power rating increases rapidly with increasing load mismatch which makes them very expensive.

The joint calibration scheme for synchronizing the voltage and current probes introduced in this work was developed in connection with the authors' work on APPJs being undertaken in their laboratory. The procedure eliminates the problem of phase errors that plague V-I probes when these are operated as independent probes *without* joint calibration. The method relies on designing a set-up, wherein *both* the *true voltage and true current* can be determined *independently* of the V and I probes, so that comparing the *known* voltage and current with those measured by the V and I probes yields *new calibration constants* for the two probes. Using these new calibration constants allows one to obtain the correct voltage and current fed to the APPJ from the voltages and currents *measured* by the probes.

For a full resolution of the problem, one needs to be able to determine the *magnitudes and phases* of the complex signal amplitudes acquired by the oscilloscope. This work was accomplished using a special FFT algorithm capable of minimizing aliasing errors so that the phases and magnitudes of the captured signals are resolved accurately. Following FFT analysis, the data may be used to estimate different physical quantities of interest like the average power absorbed by the plasma ($P_{av}$), the complex plasma impedance ($Z_p$), power in the harmonics produced by the plasma, forward and reflected wave amplitudes, etc.

The paper is organized as follows. The joint calibration technique is discussed in Section II. Post-calibration, data acquired on the APPJ for which the procedure was developed are presented in Section III to illustrate the efficacy of the technique. A discussion of the FFT procedure used for eliminating aliasing errors is also given there. For certain applications or situations, the APPJ device may have to be placed remotely from the V-I probes. In such situations, suitable voltage and current transformation formulas (derived in the Appendix) have to be used to convert the signal values measured at the location of the probes to those at the APPJ terminals.

## II. JOINT CALIBRATION OF V-I PROBES

*Joint calibration procedure:* As noted above, commercially procured voltage and current probes are most often only provided with the signal magnitude or scale conversion data. Thus no time or phase delay corrections *between the probe output read on the oscilloscope and the true signal value at the location of the probe* are provided. Thus, at high frequencies, it is eminently possible for time or phase errors to arise in high bandwidth probes due to variations in the time delays in probe electronics and the electrical lengths of the cables with the frequency. It will be seen that the problem can be eliminated by a *joint calibration* of the V-I probes. As noted earlier, the calibration procedure involves generating a set of new calibration constants for the two probes from which one may determine the true voltage, $V_t$ and the true current, $I_t$ ($V_t$ and $I_t$ being complex, RMS amplitudes). For mounting the probes and ensuring that they are placed as close to each other as possible, a small and compact, shielded metallic box-type *adapter* was fabricated. The box is provided with N-type connectors and can be inserted in series with the device to be tested. The V and I probes may be inserted into the box and connected robustly to pick up the voltage and current on the transmission line.

The calibration procedure involves connecting a signal generator (Agilent N9310A) to an oscilloscope (Lecroy Teledyne Model HD 4096, 12-bit) directly via the adapter box *without any intervening cables* (see **Figure 1**). The latter ensures that the length of the connecting line between the signal generator and the oscilloscope is as short as possible (~ few cm) giving negligible time delay or phase shift between the actual signal on the line (within the box) and that measured at the oscilloscope. Thus the voltage recorded on the oscilloscope may be regarded as the *true voltage signal $V_t$ on the line within the box*. The voltage probe (Tektronix - Model P6015A, 75 MHz,) and the current probe (Lecroy - Model CP031A, 100 MHz,) are also connected to other channels on the *same* oscilloscope via their respective connecting cables. The figure also shows that the V-probe impedance $Z_V$ appears as shunt impedance between the line and the ground, whereas the I-probe impedance $Z_I$ appears in a series in the line. The values of $Z_V$ and $Z_I$ are provided by the manufacturers and are given in **Figure 1**. It is apparent that $Z_V$ appears in parallel with the input impedance $Z_O$ of the oscilloscope, while the impedance of the current probe may be neglected as $|Z_I|$ ~ few milliohms. Other impedances $Z_d$ were also connected as shown in **Figure 1**. Typically, $Z_d$ may be chosen to be a suitable combination of capacitors and resistors.

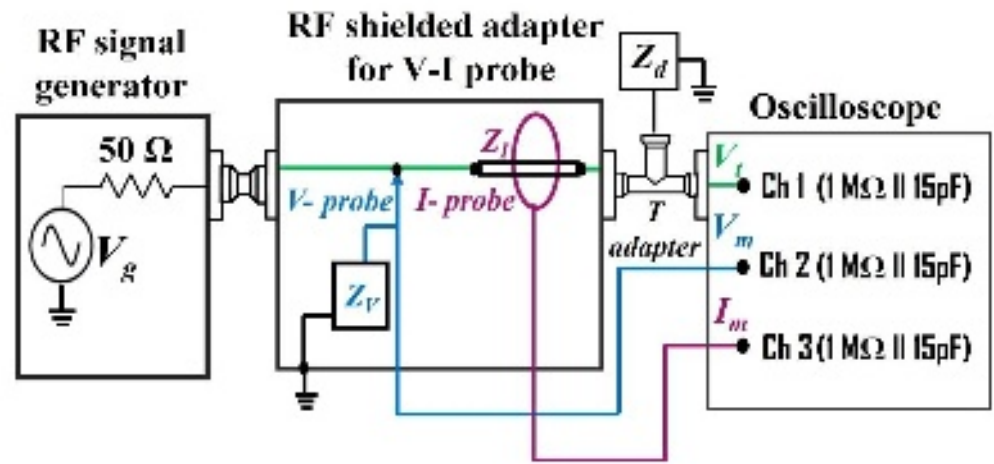

**Figure 1: The scheme for V-I probe joint calibration, where, $Z_V$ = 100 MΩ || 3pF, $|Z_I|$ = 10 mΩ, $Z_d$ = 50 Ω + $j\omega C_d$, $C_d$ = 15 - 83pF, $Z_o$ = 1 MΩ || 15pF**

The voltage signal, $V_m$ from the V probe (as given by the oscilloscope) is labelled the *measured voltage signal*. It turns out that the two signals (measured and true) are not identical, particularly for frequencies in the tens of MHz range. The ratio, $C_V = \frac{V_m}{V_t}$ gives the *voltage calibration constant*. In normal use, dividing the voltage measured using the V probe by $C_V$ would yield the true voltage on the line. Likewise, one needs to determine the true current $I_t$ on the line first. One derives $I_t$ from $V_t$ as follows: The total impedance at the oscilloscope terminal is the oscilloscope impedance ($=Z_O$) in parallel with $Z_V$ (the impedance of the V probe) and the load impedance $Z_d$. As noted above, the impedance of the current probe (~ milliohms) comes in series and is neglected (though it is not necessary to do so). Letting, $Z_t = Z_o \parallel Z_V \parallel Z_d$ one obtains, $I_t = \frac{V_t}{Z_t}$ as the *true current* on the transmission line. *It is this latter procedure of* connecting $V_t$ and $I_t$ that synchronizes the V- and I-probes and eliminates phase and amplitude errors in each probe. Analogous to the voltage, one defines a *current calibration constant*, $C_I = \frac{I_m}{I_t}$, where $I_m$ is the current *measured* using the I-probe. For determining an unknown current, one obtains the true current by dividing the current measured using I-probe by $C_I$. To lend robustness to the calibration, $Z_d$ was constructed using different capacitors (15, 27, 47 and 83 pF) in series with a 50 Ω resistor. Additional precautions like keeping the number of connectors between the adapter box, the load, the RF generator and the oscilloscope to a minimum were also taken. The voltage and current joint calibration constants of the V and I probes averaged over several values of the complex test load impedance $Z_d$, with each experiment repeated five times, are as follows.

$$C_V = 0.90 \angle -77.35° \text{ and } C_I = 1.47 \angle -71.65° \tag{1}$$

It may be kept in mind that the joint calibration process has to be carried out *separately* for each frequency component of interest (the fundamental and relevant harmonics) for the plasma device at hand. In the present work results for only the fundamental are presented. Finally, given the measured voltage and current $V_m$ and $I_m$ the true voltage and current ($V_t$ and $I_t$) are given by,

$$V_t = \frac{V_m}{C_V} \quad \text{and} \quad I_t = \frac{I_m}{C_I} \tag{2}$$

***Discussion on the validity of calibration procedure:*** V-I probes are important diagnostic tools for APPJs since they provide an *in-depth electrical characterization* of APPJs. Consequently, the synchronized calibration of the *two complex* (or four real) variables generated by V-I probe measurements is an indispensable step in the use of V-I probes for APPJs.

In this context, it would be useful to have clarity on the following issue first: For measurements conducted using a *different* pair of V-I probes for *each* measurement (chosen from an ensemble of V-I probes), the measured phases and amplitudes would be *randomly distributed* and the associated error would be a random error. On the other hand, *repeated* measurements using the *same or a given set* of V-I probes would yield phases and amplitudes that would have a systematic deviation from their actual values and the associated error would be a systematic error. The latter is essentially an accuracy issue, which has to be resolved by a suitable calibration procedure. Any noise added due to random fluctuations in the plasma has to be eliminated by averaging. It is precisely this that has been attempted in the present work.

A second issue has to do with the notion of calibration, which implies comparison with a reference. In the present case, for instance, one has to generate a reference for the complex voltage and current being measured by the V-I probes. *The latter implies that one has also to measure the voltage and current at the location of the probes by an independent method such that it is apparent from the measurement technique that the voltage and current being measured by this method are actually the 'true' voltage and current at that point*. If this is assured, one may assert that the proposed *'methodology'* provides satisfactory 'references' for calibrating the V-I probes.

In light of the above, the most important step is to *design a setup that allows one to determine* the 'true' voltage and 'true' current *on the line at the location of the probes, completely independently of the probes*. One may label the voltage and current measured by the V-I probes as, $V_m$ and $I_m$ and the 'true' voltage and current determined independently of the probes as, $V_t$ and $I_t$. It is obvious that the design of the set-up should be such that the validity or correctness of the calibration process is self-evident. As already seen above, the previous subsection describes precisely how such a set-up was fabricated and used. Also, the design of the set-up makes it apparent at once, without any room for any doubt, that the signals measured using the set up are indeed the 'true' voltage and current ($V_t$ and $I_t$), from which the required calibration constants may be determined.

***FFT analysis of signals recorded on the oscilloscope:*** The phases and magnitudes of the complex voltage and current signal data ($V_t$, $V_m$ and $I_m$) collected on the oscilloscope were extracted using FFT. However, to do so accurately one has to eliminate aliasing errors that arise during FFT analysis. In the present work, the special algorithm used for eliminating such errors is based on the authors' previous work on the use of dual-directional couplers for capacitively coupled plasmas at radio frequencies [29]. Details of the FFT procedure for eliminating aliasing errors are presented in the cited paper [29]. As discussed there, suppressing aliasing errors requires a particular number of samples to be used in the data sets for a *given sampling rate* (= 2.5 GS/s) *and signal frequency* (= 13.56 MHz).

Figure 2 gives the complex FFT amplitudes (magnitudes and phases) without correction (red data points) and with correction (black data points) for aliasing for a typical data set in the present work. In the first case, as seen in **Figure 2 (a),** one simply chooses the total number of data points in the sample as collected (= 2502), which yields the signal peak not at 13.56 MHz but 13.98 MHz, with a height of ~ 242.6 V. One also sees a distortion in the corresponding phase plot (**Figure 2 (b)).** On the other hand, as explained in Ref 29, choosing the number of data points as per the FFT procedure for eliminating aliasing one uses only 2028 data points in a single sample. Both the magnitude and phase plots exhibit the peak at 13.56 MHz as expected with a higher signal magnitude (~ 329.2 V).

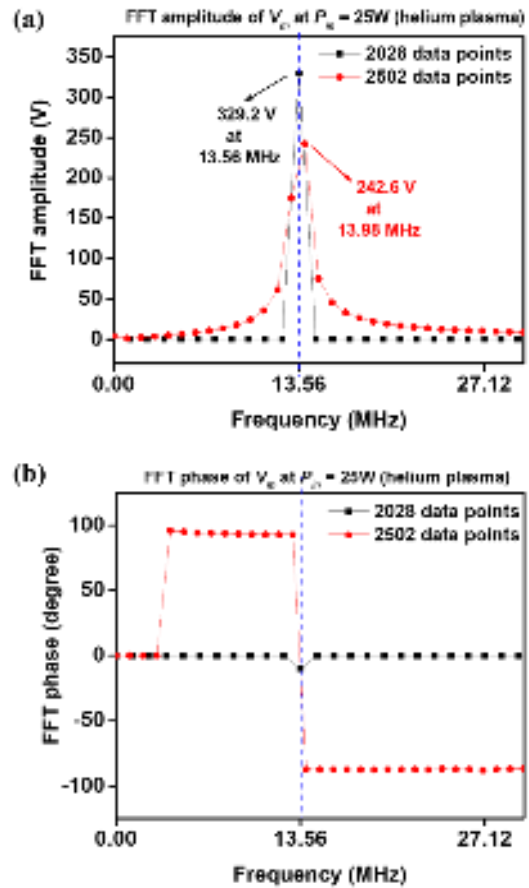

**Figure 2: FFT spectrum of a typically measured voltage ($V_m$) obtained for helium APPJ at ~ 25 W RF input power, showing errors due to aliasing: without correction (red data points) and with correction (black data points). (a) magnitude spectrum (b) phase spectrum.**

## III. RESULTS AND DISCUSSION

***The APPJ device:*** Following joint calibration of the V-I probes, they were used on the RF APPJ device being used in the laboratory. The set-up including the RF circuit and the placement of the probes is shown in **Figure 3** and is similar to those reported elsewhere [3,4]. The device itself is a custom-designed Pyrex glass tube of ≈ 5 mm outer diameter, ≈ 3 mm inner diameter and ≈ 10 cm length, with a cross-feed gas port. A copper electrode (dia. ≈ 1.5 mm) is introduced coaxially into the glass tube (up to ≈ 3 mm above its opening) and a copper strip of ≈ 3 mm width, wrapped around the tube just above the nozzle, is used as the ground electrode. Applying RF power to the central electrode along with gas (argon or helium) fed to the device ignites the discharge, which appears as a plasma jet in the ambient air. Power is applied from a 1kW, RF generator (CESAR 1310) at 13.56 MHz via its associated automatic matching network (MN) and a customized, LC series resonance circuit ($L = 1.95$ µH; Variable vacuum capacitance, $C = 12 - 500$ pF; Model: CVDD-500 Jennings). The latter makes plasma generation possible at low RF power (~2 W) in helium gas.

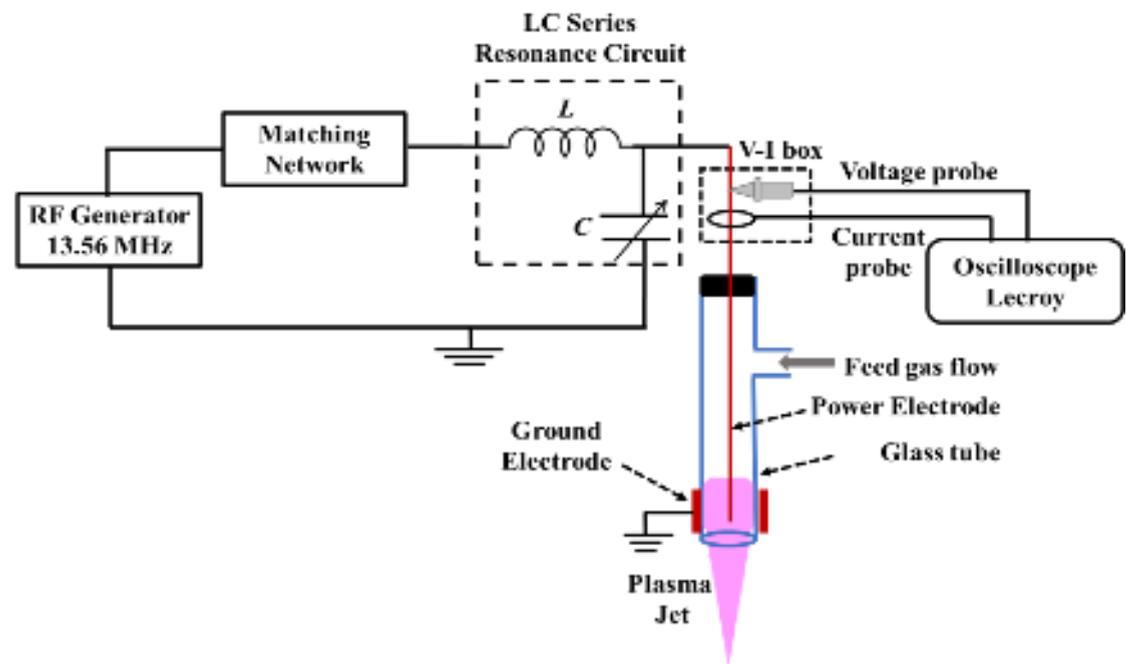

**Figure 3: Schematic of atmospheric pressure RF plasma jet showing the experimental setup and the placement of V-I probes**

As discussed earlier, the impedance of the small plasma device, the APPJ, is predominantly capacitive, arising from a series combination of the plasma sheath capacitance and the capacitance of the glass wall of the tube. Consequently, the phase difference between the voltage and current signals is very nearly 90°, ($\cong 86° - 88°$; voltage lagging current). It may be kept in mind that power absorption is proportional to the *resistive* part of the impedance, which is given by the cosine of the phase difference and becomes a highly sensitive function of the phase difference when the latter is very close to 90°. Thus it is important to be able to resolve the phase difference accurately to ensure the accuracy and robustness of the probe measurements and eliminate errors as much as possible.

***Verification of joint calibration efficacy:*** The effectiveness of the joint calibration procedure for helium APPJ is summarized in Table 1, where the complex, RMS amplitudes $V_m$, $I_m$, $V_t$, $I_t$ (related by Eqn. (2)) are given along with the phase differences, $\varphi_m = \angle V_m - \angle I_m$ and $\varphi_t = \angle V_t - \angle I_t$, for different RF input power, $P_{in}$. It may be noted that $V_m$ and $I_m$ presented in Table 1 were obtained by averaging the signals (over 250 data sets) using the averaging function of the oscilloscope to eliminate the jitter in the signals due to noise from the plasma (see subsection on Precautions in section III). It is seen from the Table that $|\varphi_m| > 90°$ in all cases, indicating that there is a genuine problem that persists even after averaging. On the other hand, the phase angle $\varphi_t = \angle V_t - \angle I_t$ obtained *after* calibration is seen to be free of this anomaly in all cases; it is seen that $\varphi_t$, though close to - 90º, *never* crosses it. It must be remembered that during measurements, although the voltage across the plasma load (APPJ) is, $V_L = V_t$, the current through the load is, $I_L = I_t - V_t/Z_V$, since $Z_V$, the V-probe impedance comes in parallel and has to be subtracted. However, because $|Z_V|$ is very large ($\gg 1$), it turns out $I_L \approx I_t$ and so for sake of simplicity, we shall regard $V_t$ and $I_t$ as the load voltage and load current. Thus, the average power fed to the plasma, $P_{av} \cong |V_T| \cdot |I_T| \cos(\varphi_t)$.

The $P_{av}$ fed to the plasma load is also computed for different $P_{in}$ and given in **Table 1**. It can be seen that there is a considerable disparity between the values of $P_{av}$ and $P_{in}$. The reason for this has to do with the fact that in the present experiments only a few watts of RF power were being drawn from a ~1000 W RF generator. In such a high power generator, the capacity to resolve small forward and reflected power levels is limited. Normally in such situations, a dual directional coupler (DDC) would be required between the matching network (MN) and the RF generator (along with a power sensor) to measure both forward and reflected power accurately. In the present experiments, however, a DDC was not available and so one had to depend on the forward and reflected power readouts on the generator for impedance matching and tuning. It is believed that the inability of the high-power generator to resolve low reflected powers (typically below several watts) resulted erroneously in its indicating a matched condition when actually true matching had not been achieved. In high power operation (~ few hundred watts) an error of a few tens of watts in both forward and reflected power would not be significant, whereas in the present case, where only a few watts were involved, the discrepancies in forward and reflected powers could be significant. In **Table 1** and in what follows, the $P_{in}$ values serve simply as an indicator of the input power levels and the corresponding values of $P_{av}$, as the *true* measure of the actual power fed into the plasma load.

Table 1: $V_t$ (= $V_L$), $I_t$ ( ≈ $I_L$), $\varphi_t = \angle V_t - \angle I_t$ and $P_{av}$ obtained for He APPJ after joint calibration of the V-I probes for various input power, $P_{in}$; $V_m$ and $I_m$ are the voltages and currents measured using the V-I probes after averaging; $\varphi_m = \angle V_m - \angle I_m$.

| $P_{in}$ (W) | $V_m$ (V) | $I_m$ (A) | $\varphi_m$ | $V_t$ (V) | $I_t$ (A) | $\varphi_t$ | $P_{av}$ (W) |
|---|---|---|---|---|---|---|---|
| 10 | 221.5∠10.3° | 0.33 ∠81.6° | - 91.9° | 247.4 ∠67.1° | 0.22 ∠153.3° | - 86.2° | 3.6 |
| 15 | 263.4∠11.0° | 0.39 ∠80.7° | - 91.7° | 294.5 ∠66.3° | 0.27 ∠152.3° | - 86.0° | 5.5 |
| 20 | 298.8∠10.6° | 0.45 ∠80.8° | - 91.4° | 333.9 ∠66.7° | 0.31 ∠152.5° | - 85.8° | 7.6 |
| 25 | 329.2∠10.1° | 0.51 ∠81.1° | - 91.2° | 367.8 ∠67.2° | 0.34 ∠152.7° | - 85.5° | 9.8 |

***Transformation of $V_t$ and $I_t$ to $V_p$ and $I_p$ at APPJ terminals:*** For certain applications and situations it is possible that the V-I probes, though placed after the MN, may not be located close to the APPJ. Under such circumstances, the transmission line connecting the probes to the APPJ would add delays (phase shifts) to the RF voltage and current and might also result in their attenuation due to losses in the line. These could result in modification of the input plasma impedance and as well as a decrease in $P_{av}$. It is, therefore, *necessary to be able to determine the* *voltage and current at the APPJ terminals,* for determining the actual power delivered to the APPJ, its input impedance, etc. The transformation equations connecting the voltage and current $(V_p, I_p)$ at the APPJ electrodes to the load / true voltage and current $(V_t, I_t)$ at the measurement plane at the location of the probes is given as:

$$V_p = \frac{V_t}{1+\rho_V} e^{-j\beta w}\left(1 + \rho_V e^{2j\beta w}\right) \tag{3}$$

$$I_p = \frac{V_t}{Z_0(1+\rho_V)} e^{-j\beta w}\left(1 - \rho_V e^{2j\beta w}\right) \tag{4}$$

In Eqns. (3) and (4), $Z_0$ and $\beta$ are respectively, the characteristic impedance and propagation constant of the transmission line between the measurement plane and the APPJ; $\rho_V$ $(= \frac{Z_t - Z_0}{Z_t + Z_0})$ and $Z_t = \frac{V_t}{I_t}$ are the voltage reflection coefficient and the input impedance *at the measurement plane* and *w* is the length of the line. The derivation of Eqns. (3) and (4) is given in the **Appendix**. In the derivation, the line has been assumed to be lossless for simplicity. Knowledge of $V_p$ and $I_p$ yields both the power delivered to the discharge, $P_{av} = |V_p| \cdot |I_p| \cos(\angle V_p - \angle I_p)$ as well as the complex impedance, $Z_p = \frac{|V_P|}{|I_P|}(\angle V_p - \angle I_p)$ *at the APPJ terminals*.

In the present work, the connection from the RF input (at the measurement plane) to the APPJ terminals may be represented as a short, *two-conductor transmission line* with a phase constant $\beta = 0.28$ rad/m, and line length, $w = 10$ cm. Since the line is short in comparison to the wavelength on the line, the changes in the voltage and current are small. Using the voltage and current $(V_t, I_t)$ data and using Eqns. (3) and (4), and the definitions are given above, one may determine, $P_{av}$ the net power fed (absorbed) to (by) the plasma and $Z_p$, the input impedance at the APPJ terminals for helium and argon APPJs for various RF input power $P_{in}$. The results are summarized in **Table 2**.

**Table 2: Average power and input impedance of plasma discharge at APPJ terminals for various input power**

| $P_{in}$ (W) | Helium plasma | | Argon plasma | |
|---|---|---|---|---|
| | $P_{av}$ (W) | $Z_p$ (Ω) | $P_{av}$ (W) | $Z_p$ (Ω) |
| 10 | 3.6 | 80.2- j 1172 | 2.5 | 58.6- j 1195 |
| 15 | 5.5 | 84.6- j 1161 | 4.5 | 75.9- j 1194 |
| 20 | 7.6 | 89.8- j 1150 | 5.9 | 79.6- j 1187 |
| 25 | 9.8 | 93.6- j 1139 | 7.7 | 81.5- j 1173 |

Owing to the series combination of the sheath and glass wall capacitances, $Z_p$ is predominantly capacitive. The resistive part of $Z_p$, given by $R_p = \text{Re}(Z_p) = P_{av}/|I_p|^2$ gives the *overall plasma resistance that includes all losses*. It is to be distinguished from the resistance that one computes from the circuit model of the plasma, which does not include power losses due to plasma particle loss, gas heating, etc. It is seen that $R_p$ increases with $P_{av}$; this is as it should be since both power loss in the plasma circuit and that due to escaping particles increase with $P_{av}$.

***Precautions to be observed during V-I probe measurements at radio frequencies:*** In order to carry out accurate RF measurements, a few precautions to be observed are discussed here.

**(i)** It is crucial to be able to determine accurately the complex voltage and current amplitudes measured by the V and I probe both for the fundamental as well as any significant harmonics produced by the plasma. Though in principle it should be a simple matter to do so, in actual practice it is possible for phase errors to arise (particularly for high bandwidth probes) due to variations in the time delay in the probe electronics and uncertainties in the electrical lengths of cables with the frequency. In the present work, this is taken care of by the joint calibration technique. The latter not only eliminates the inaccuracies of the individual probes but also ensures that the two probes are synchronized. To be able to do this, a primary requirement is to set up a test system in which the *true voltage and true current are known a priori independently* of the V and I probes so that comparing the latter with those measured by the V and I probes yields *new calibration constants* for the two probes.

**(ii)** In standard RF cables, typically a cable length of ~ 4.5 cm gives a phase delay ≈ 1° at a frequency of ≈ 13.56 MHz. This means that to determine the phases measured by the V and I probes accurately, the phase shifts generated by the gap between the probes (when placed inside the adapter box) and/or caused by any other path differences should also be accounted for by adding (subtracting) it to (from) the overall phase shift.

**(iii)** It may happen that for ease of end-use or a remote application site, one may have to employ long RF cables between the V and I probes and APPJ. In the latter situation, one needs to use appropriate transformation formulas (given in the previous section) to obtain the voltage and current at the APPJ terminals accurately, both for the fundamental as well as for the harmonics generated by the plasma.

**(iv)** To ensure high accuracy of signal measurement and resolution, it helps to have a 12-bit (4096:1 resolution) or higher, digital storage oscilloscope with a high sampling rate (2.5 GS/s). On the other hand, to minimize the jitter, the waveforms may be acquired by operating the oscilloscope in averaging mode (typically ~256 measurements).

**(v)** Further, the complex amplitudes (magnitude and phase) of all signals need to be carefully analyzed using suitable FFT techniques that eliminate aliasing errors.

## IV. CONCLUSIONS

To conclude, this paper introduces for the first time a joint calibration scheme for synchronizing voltage and current probes that can enhance the accuracy of the probe measurements significantly and would be particularly useful for RF-based plasma systems, especially for APPJs. Using actual experimental results from measurements on RF APPJs, it is shown that the calibration technique introduced is able to resolve phase shifts very close to 90°, but without exceeding it. Further, for accurate determination of the phases and magnitudes of the complex signal amplitudes, the latter are analyzed using special FFT algorithms capable of minimizing/eliminating aliasing errors. The data is also used to determine the average RF power fed to the device and its complex plasma impedance. With the joint calibration procedure in place, V-I probes can indeed be an accurate and economical diagnostic tool for RF plasma devices like APPJs, since the latter are difficult to diagnose due to their small size. In particular, it is the only method that can offer important *electrical* information about the device, like the average RF power fed to the device, its complex plasma impedance, nature of the harmonic spectrum, etc. Such data can be used effectively in conjunction with a suitable plasma model for estimating other important discharge parameters of APPJ discharges with considerable sensitivity to the applied RF power, gas flow rate, etc., which is the future scope of the present work.

**Acknowledgments**

The authors wish to thank the referee for his comments and suggestions that have improved the paper substantially.

**Appendix**

**Transformation of $V_t$ and $I_t$ to $V_p$ and $I_p$ at APPJ terminals**

One may transform the *load / true* voltage and current, $V_t$, $I_t$ determined at the *measurement or probe plane*, $M - M'$ located at $x = 0$, to the voltage and current, $V_p$, $I_p$ at the *RF electrodes at the plasma device or* the *plasma plane*, $P - P'$ located at $x = w$, by regarding the intervening connection between the RF terminals at $M - M'$ and $P - P'$ as a *transmission line* with propagation constant $\beta$, characteristic impedance, $Z_0$ and length $l$. For radian frequency $\omega$ and harmonic time variations of the form, $\sim e^{+j\omega t}$, the forward and reflected waves will vary as, $e^{-j\beta x}$ and $e^{+j\beta x}$, respectively.

Let $V^+(x) = V_0^+ \, e^{-j\beta x}$ and $V^-(x) = V_0^- \, e^{+j\beta x}$ represent respectively, the complex, forward and reflected voltage wave amplitudes on the transmission line at a location $x$; $V_0^+$ and $V_0^-$ are forward and reflected wave amplitudes at $x = 0$ or at $M - M'$. The total voltage on the line at $x$ equals, $V(x) = V^+(x) + V^-(x)$. Likewise, one has for the current: $I^+(x) = I_0^+ \, e^{-j\beta x}$ and $I^-(x) = I_0^- \, e^{+j\beta x}$, with a total current $I(x) = I^+(x) - I^-(x)$ and, $Z_0 = \frac{V_0^+}{I_0^+} = \frac{V_0^-}{I_0^-}$. Note that the reflected current has to be subtracted as it flows in the reverse direction. It may be noted that the voltage and current at $M - M'$ or at $x = 0$, may be written as: $V_t = V(0) = V_0^+ + V_0^-$, and $I_t = I(0) = I_0^+ - I_0^-$. The impedance and the voltage reflection coefficient at $x = 0$ are respectively, $Z_t = Z(0) = \frac{V_t}{I_t}$ and $\rho_V = \frac{V_0^-}{V_0^+}$. Using the given relations and definitions, one may solve for $V_0^+$ and $V_0^-$ from $V_t, I_t$, giving:

$$V_0^+ = \frac{V_t + Z_0 I_t}{2} = \frac{V_t (\bar{Z}_t + 1)}{2 \bar{Z}_t} \tag{A-1}$$

$$V_0^- = \frac{V_t - Z_0 I_t}{2} = \frac{V_t (\bar{Z}_t - 1)}{2 \bar{Z}_t} \tag{A-2}$$

with $\bar{Z}_t = \frac{Z_t}{Z_0} = \frac{1 + \rho_V}{1 - \rho_V}$. Using the above definitions, it is straightforward to derive $V_p$ and $I_p$ at $P - P'$ or at $x = w$:

$$V_p = V_0^+ e^{-j\beta w}\{1 + \rho_V e^{2j\beta w}\} = \left(\frac{V_t}{1 + \rho_V}\right) e^{-j\beta w}\{1 + \rho_V e^{2j\beta w}\} \tag{A-3}$$

$$I_p = \frac{V_0^+}{Z_0} e^{-j\beta w}\{1 - \rho_V e^{2j\beta w}\} = \frac{V_t}{Z_0(1 + \rho_V)} e^{-j\beta w}\{1 - \rho_V e^{2j\beta w}\} \tag{A-4}$$

In the present derivation, the line has been assumed to be lossless for simplicity, which is a reasonable assumption for moderately long, low loss lines. However, it is straightforward to extend the above to include loss in the line. One simply has, $j\beta \rightarrow \alpha + j\beta$, where $\alpha$ gives the attenuation coefficient of the line in Np / m.

**References**

[1] I.E. Kieft, E.P.V.D. Laan, and E. Stoffels, New J. Phys. **6**, 1 (2004).

[2] M.A. Sobolewski, J. Vac. Sci. Technol. A Vacuum, Surfaces, Film. **10**, 3550 (1992).

[3] J. Golda, J. Held, B. Redeker, M. Konkowski, P. Beijer, A. Sobota, G. Kroesen, N.S.J. Braithwaite, S. Reuter, M.M. Turner, T. Gans, D. O'Connell, and V. Schulz-Von Der Gathen, J. Phys. D. Appl. Phys. **49**, 084003 (2016).

[4] D. Gahan and M.B. Hopkins, J. Appl. Phys. **100**, (2006).

[5] B. Bora, H. Bhuyan, M. Favre, E. Wyndham, H. Chuaqui, and C.S. Wong, Curr. Appl. Phys. **13**, 1448 (2013).

[6] S. Hofmann, A.F.H. Van Gessel, T. Verreycken, and P. Bruggeman, Plasma Sources Sci. Technol. **20**, (2011).

[7] D. Marinov and N.S.J. Braithwaite, Plasma Sources Sci. Technol. **23**, 62005 (2014).

[8] J. Golda, F. Kogelheide, P. Awakowicz, and V.S. Der Gathen, Plasma Sources Sci. Technol. **28**, 95023 (2019).

[9] J. Golda, J. Held, and V.S. Von Der Gathen, Plasma Sources Sci. Technol. **29**, 025014 (2020).

[10] V. Léveillé and S. Coulombe, Meas. Sci. Technol. **17**, 3027 (2006).

[11] B. Bora, H. Bhuyan, M. Favre, H. Chuaqui, E. Wyndham, and M. Kakati, Phys. Plasmas **19**, 1 (2012).

[12] P.A.C. Beijer, A. Sobota, E.M. Van Veldhuizen, and G.M.W. Kroesen, J. Phys. D. Appl. Phys. **49**, 104001 (2016).

[13] N. Spiliopoulos, D. Mataras, and D.E. Rapakoulias, J. Vac. Sci. Technol. A Vacuum, Surfaces, Film. **14**, 2757 (1996).

[14] L. Ge and Y. Zhang, Plasma Sci. Technol. **16**, 924 (2014).

[15] G.V. Prakash, N. Behera, K. Patel, and A. Kumar, AIP Adv. **11**, 085329 (2021).

[16] A. Shashurin and M. Keidar, Phys. Plasmas **22**, (2015).

[17] A.F.H. Van Gessel, B. Hrycak, M. Jasiński, J. Mizeraczyk, J.J.A.M. Van Der Mullen, and P.J. Bruggeman, J. Phys. D. Appl. Phys. **46**, (2013).

[18] A.F.H. Van Gessel, E.A.D. Carbone, P.J. Bruggeman, and J.J.A.M. Van Der Mullen, Plasma Sources Sci. Technol. **21**, (2012).

[19] S. Hubner, J.S. Sousa, J. Van Der Mullen, and W.G. Graham, Plasma Sources Sci. Technol. **24**, (2015).

[20] S. Zhang, A. Sobota, E.M. Van Veldhuizen, and P.J. Bruggeman, J. Phys. D. Appl. Phys. **48**, 015203 (2014).

[21] J. Jiang, V.S.S.K. Kondeti, G. Nayak, and P.J. Bruggeman, J. Phys. D. Appl. Phys. **55**, 225206 (2022).

[22] J. Jiang and P.J. Bruggeman, Plasma Sources Sci. Technol. **30**, 105007 (2021).

[23] D. Gidon, D.B. Graves, A. Mesbah -, Z. Liu, D. Xu, D. Liu, al -, J. Jiang, and P.J. Bruggeman, J. Phys. D. Appl. Phys. **54**, 15LT01 (2021).

[24] J. Park, I. Henins, H.W. Herrmann, G.S. Selwyn, and R.F. Hicks, J. Appl. Phys. **89**, 20 (2001).

[25] A.D. MacDonald and S.C. Brown, Phys. Rev. **75**, 411 (1949).

[26] Mahreen, G.V. Prakash, S. Kar, D. Sahu, and A. Ganguli, J. Appl. Phys. **130**, 083301 (2021).

[27] Mahreen, G.V. Prakash, S. Kar, D. Sahu, and A. Ganguli, Contrib. to Plasma Phys. **62**, e202200007 (2022).

[28] S. Das, V.P. Gajula, S. Mohapatra, G. Singh, and S. Kar, Heal. Sci. Rev. **4**, 100037 (2022).

[29] M. Dünnbier, M.M. Becker, S. Iseni, R. Bansemer, D. Loffhagen, S. Reuter, and K.D. Weltmann, Plasma Sources Sci. Technol. **24**, 65018 (2015).

[30] A. Rawat, A. Ganguli, R. Narayanan, and R.D. Tarey, Rev. Sci. Instrum. **91**, 094705 (2020).

[25] A.D. MacDonald and S.C. Brown, Phys. Rev. **75**, 411 (1949).

[26] Mahreen, G.V. Prakash, S. Kar, D. Sahu, and A. Ganguli, J. Appl. Phys. **130**, 083301 (2021).

[27] Mahreen, G.V. Prakash, S. Kar, D. Sahu, and A. Ganguli, Contrib. to Plasma Phys. **62**, e202200007 (2022).

[28] S. Das, V.P. Gajula, S. Mohapatra, G. Singh, and S. Kar, Heal. Sci. Rev. **4**, 100037 (2022).

[29] M. Dünnbier, M.M. Becker, S. Iseni, R. Bansemer, D. Loffhagen, S. Reuter, and K.D. Weltmann, Plasma Sources Sci. Technol. **24**, 65018 (2015).

[30] A. Rawat, A. Ganguli, R. Narayanan, and R.D. Tarey, Rev. Sci. Instrum. **91**, 094705 (2020).

18
[25] A.D. MacDonald and S.C. Brown, Phys. Rev. **75**, 411 (1949).

[26] Mahreen, G.V. Prakash, S. Kar, D. Sahu, and A. Ganguli, J. Appl. Phys. **130**, 083301 (2021).

[27] Mahreen, G.V. Prakash, S. Kar, D. Sahu, and A. Ganguli, Contrib. to Plasma Phys. **62**, e202200007 (2022).

[28] S. Das, V.P. Gajula, S. Mohapatra, G. Singh, and S. Kar, Heal. Sci. Rev. **4**, 100037 (2022).

[29] M. Dünnbier, M.M. Becker, S. Iseni, R. Bansemer, D. Loffhagen, S. Reuter, and K.D. Weltmann, Plasma Sources Sci. Technol. **24**, 65018 (2015).

[30] A. Rawat, A. Ganguli, R. Narayanan, and R.D. Tarey, Rev. Sci. Instrum. **91**, 094705 (2020).

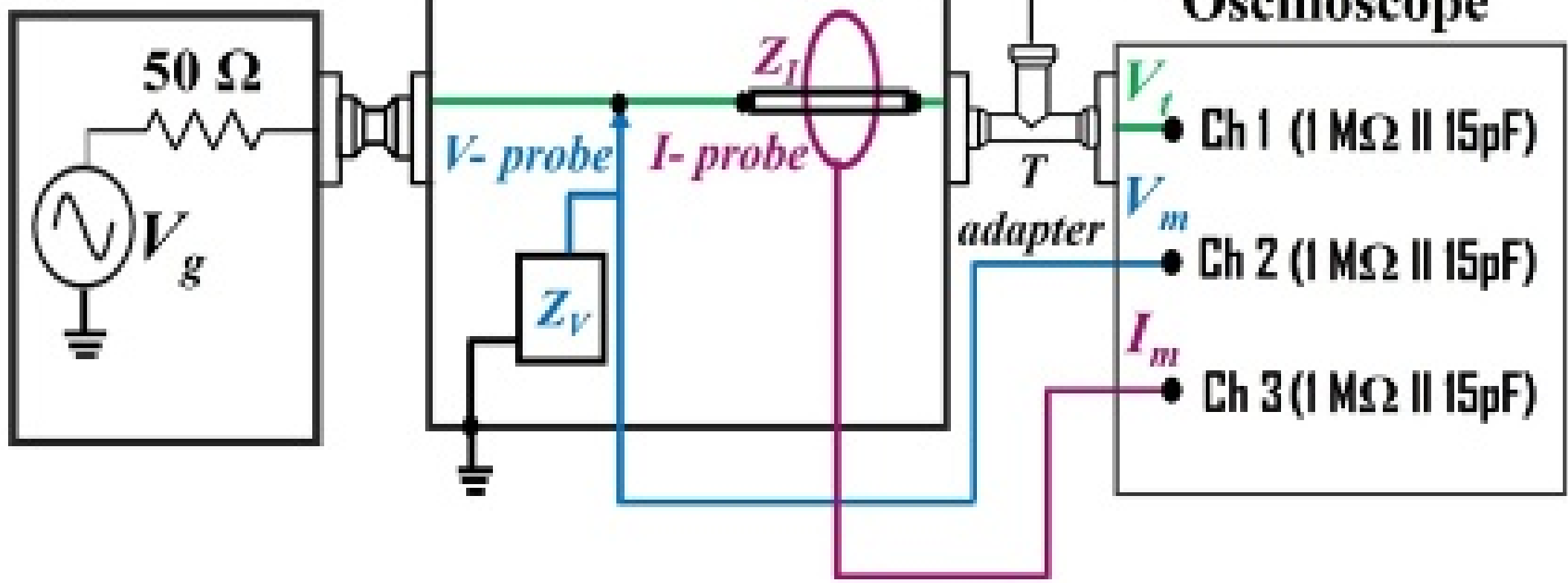

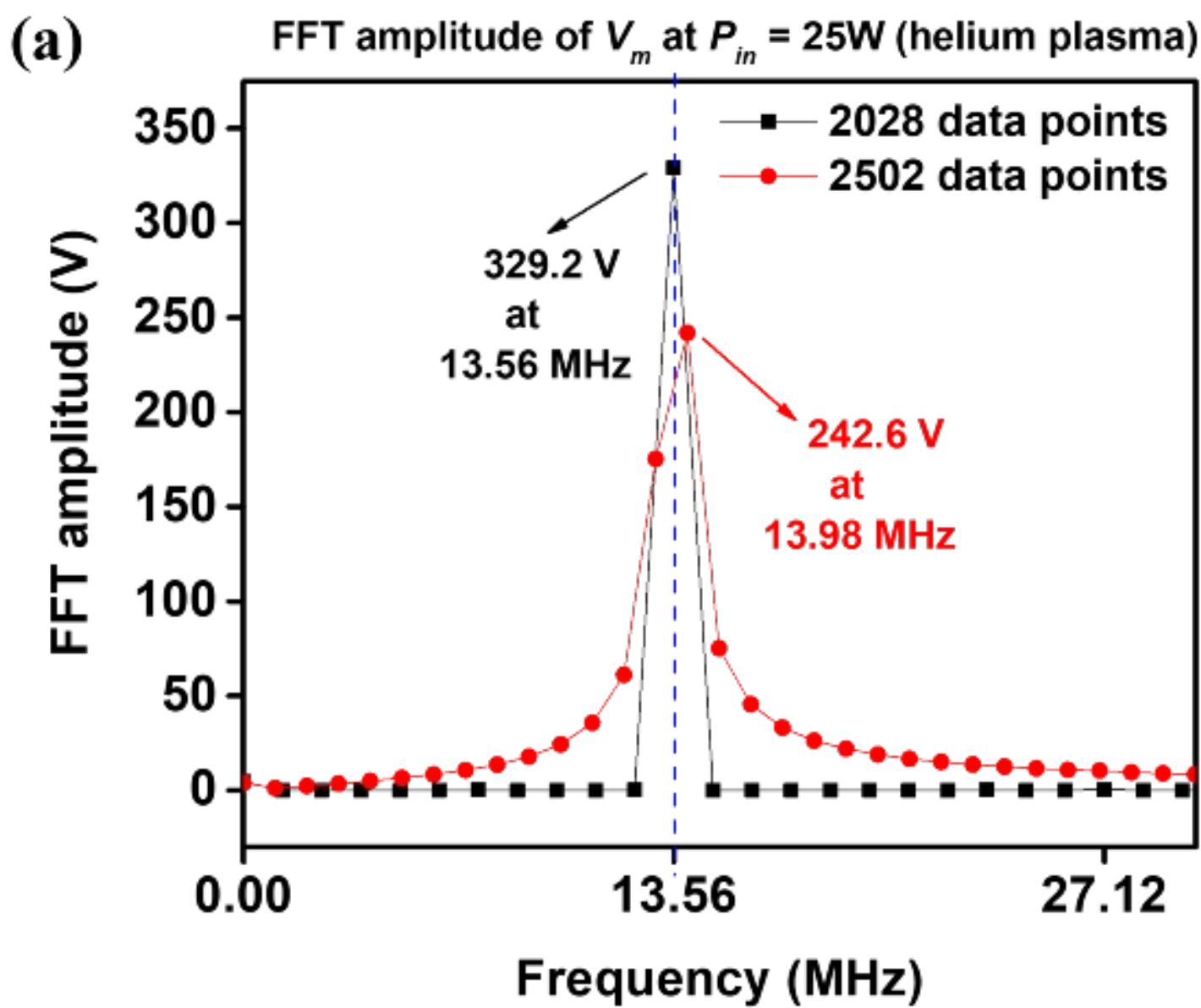

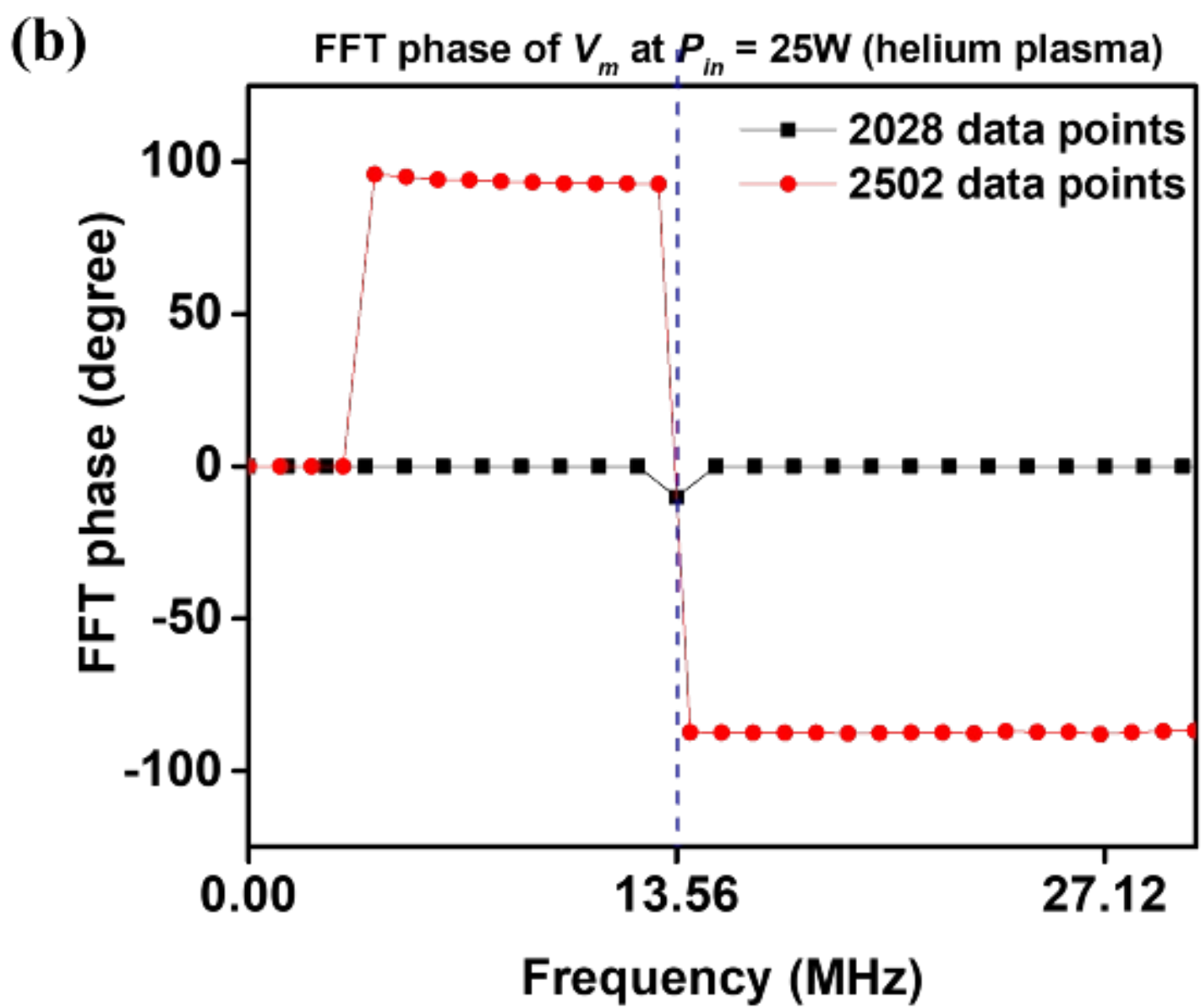

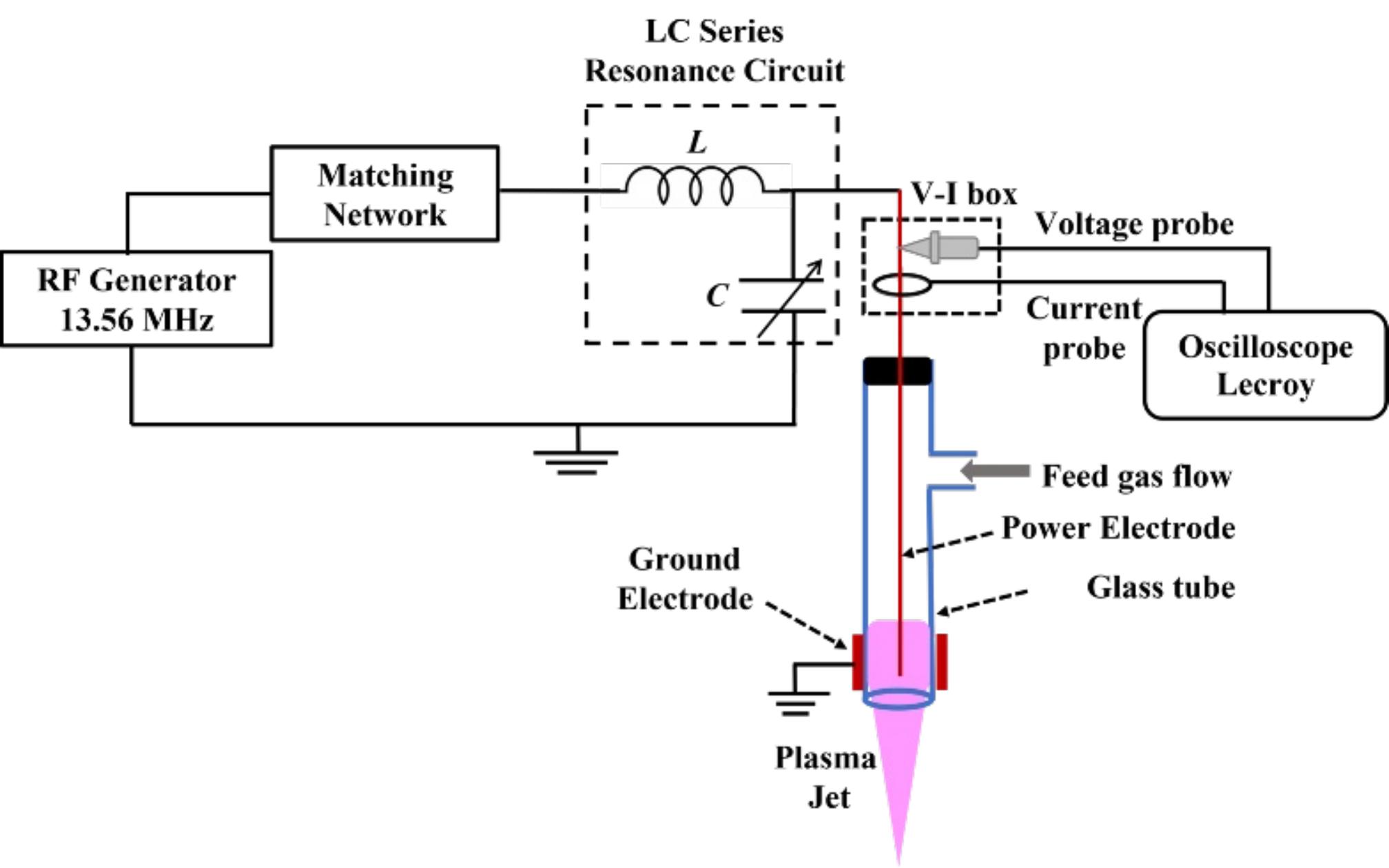